\documentclass[prb,showpacs,twocolumn,aps,superscriptaddress,floatfix]{revtex4}
\usepackage{amsmath}
\usepackage{amssymb}
\usepackage{bm}
\usepackage{graphicx}
\usepackage{color}

\begin{document}

\title{Phase separation of hydrogen atoms adsorbed on graphene and the
smoothness of the graphene-graphane interface}

\author{A.L. Rakhmanov}
\affiliation{Advanced Science Institute, RIKEN, Wako-shi, Saitama,
351-0198, Japan}
\affiliation{Institute for Theoretical and Applied Electrodynamics, Russian
Academy of Sciences, 125412 Moscow, Russia}

\author{A.V. Rozhkov}
\affiliation{Advanced Science Institute, RIKEN, Wako-shi, Saitama,
351-0198, Japan}
\affiliation{Institute for Theoretical and Applied Electrodynamics, Russian
Academy of Sciences, 125412 Moscow, Russia}

\author{A.O. Sboychakov}
\affiliation{Advanced Science Institute, RIKEN, Wako-shi, Saitama,
351-0198, Japan}
\affiliation{Institute for Theoretical and Applied Electrodynamics, Russian
Academy of Sciences, 125412 Moscow, Russia}

\author{Franco Nori}
\affiliation{Advanced Science Institute, RIKEN, Wako-shi, Saitama,
351-0198, Japan}
\affiliation{Department of Physics, University of Michigan, Ann
Arbor, MI 48109-1040, USA}

\begin{abstract}
The electronic properties of a graphene sheet with attached hydrogen atoms
is studied using a modified Falicov-Kimball model on the honeycomb lattice.
It is shown that in the ground state this system separates into two phases:
fully hydrogenated graphene (graphane) and hydrogen-free graphene. The
graphene-graphane boundary acquires a positive interface tension.
Therefore, the graphene-graphane interface becomes a straight line,
slightly rippled by thermal fluctuations. A smooth interface may be useful
for the fabrication of mesoscopic graphene-based devices.
\end{abstract}


\maketitle

\section{Introduction}

Creating a sample with flat edges is a significant challenge for producing
graphene mesoscopic
devices~\cite{meso_review}.
One possibility is to break a graphene sheet into fragments with sharp
edges.~\cite{sharp_edges} Another alternative involves the use of graphane.
Graphane~\cite{graphane_2003} is fully hydrogenated graphene; it is an
insulator with a gap of several eV. With graphane, instead of physically
cutting graphene, one can create graphene patches of required shapes inside
a sheet of graphane by local de-hydrogenation. In such systems, low-energy
electrons from graphene cannot penetrate the insulating graphane host.
Therefore, the graphene-graphane interface serves as the effective edge of
the
graphene structure. Different arrangements of this type have been
discussed.~\cite{graph,schmidt2010}
Thus, the issue of the graphene-graphane interface stability is important
both for fundamental and applied research. There are indications from
numerical studies that such interface is stable
\cite{openov_gr_interface,interface_stability2010},
and that the adsorbed hydrogens tend to cluster
together.~\cite{roman_clustering2009}
This tendency may be explained in terms of phase separation into
hydrogen-rich and hydrogen-free regions, which was established on the basis
of the semi-phenomenological analysis of the electron-mediated interaction
between hydrogen adatoms in
graphene.~\cite{shytov_separation}

The purpose of the present paper is twofold. First, we put forward a
microscopic approach to the problem of phase separation in
graphene-graphane systems. To demonstrate the phase separation 
Ref.~\onlinecite{shytov_separation}
assumed a specific type of interaction between the graphene electrons and
the adatoms. For the hydrogen adatoms this assumption is supported
experimentally and numerically. However, it remains unclear if the phase
separation is a unique feature of the hydrogen on graphene, or other
adsorbents will show the same feature. Avoiding phenomenological arguments
we discuss the phase separation within the framework of a modified
Falicov-Kimball model with an infinite interaction constant between `a
hydrogen hole' and an $s$-electron on the hydrogen atom. The advantage of
such an approach is its generality: the phase separation is a known
property of 
a ground state of
Falicov-Kimball-like
models~\cite{chaika1,chaika2,chaika3,kuras1,kuras2}
robust against variation of microscopic details. To estimate the
characteristic energies of the phase-separated state we apply the Hubbard-I
approximation.~\cite{hubbard1} 
To check the validity of this approximation, we also perform exact
diagonalization of the model Hamiltonian in a finite cluster.

The phase separation implies that the homogeneous state is either unstable
or metastable. However, it is possible to imagine that, under suitable
conditions, such phase may be stabilized for substantial amount of time. If
the stabilization is indeed possible, the properties of the homogeneous
phase can be investigated. Our calculations show that the homogeneous phase
is metallic, in agreement with the numerical results of
Ref.~\onlinecite{roman_clustering2009}.

Our second goal is to explore the connection between phase separation and
the stability of the graphene-graphane interface. We show that the
graphene-graphane interface has a positive boundary tension. To stretch the
interface with a positive interface tension by a unit length requires a
finite amount of work. This amount is high for the system considered. Thus,
the interface remains flat over substantial distances, which is a highly
desired property, necessary for the creation of ballistic mesoscopic
systems. In other words, the interface is stable not only with respect to
vacancy defects in small samples, as found in
Refs.~\onlinecite{openov_gr_interface}
and~\onlinecite{interface_stability2010},
but also with respect to any conceivable defect. Our approach allows
to obtain a qualitative estimate of the interface tension and to assess the
flatness of the interface at a given temperature. We estimate that at room
temperature the graphene-graphane interface remains atomically smooth over
distances of about $10^2$ lattice constants.

The paper is organized as follows. In
Sec.~\ref{model}
we formulate the model of the adatoms adsorbed on a graphene sample. This
model is solved in
Sec.~\ref{calcs}
within the Hubbard-I approximation. To check the accuracy of the Hubbard-I
approximation the finite-cluster numerical study is presented in 
Sec.~\ref{numerics}. In
Sec.~\ref{interface}
we investigate the stability of the graphene-graphane interface,
evaluate the interface tension, and investigate its smoothness. The
conclusions are given in
Sec.~\ref{conclusions}.

\section{Model}
\label{model}

We use the model Hamiltonian for graphane:
\begin{eqnarray}
H_{\rm A}
&=&
H_{\rm E}
-
\sum_{i \sigma}
\left[
	t_0\left(P_{i \sigma}^\dag S_{i \sigma}+ {\rm h.c.}\right)
	+
	\varepsilon_{\rm H}S_{i\sigma}^{\dag}S_{i\sigma}
\right]\!,
\label{H_A}
\\
H_{\rm E}
&=&
-\sum_{i j \sigma}
\left(
	P^\dag_{i\sigma}\hat{T}_{ij}P_{j\sigma}+ {\rm h.c.}
\right)\!,
\label{H_E}
\label{spinors}
\end{eqnarray}
where
$P^{\dag}_{i\sigma}
=\left(p^{{\cal A}\dag}_{i\sigma},p^{{\cal B}\dag}_{i\sigma}\right)$,
$S^{\dag}_{i\sigma}
=
\left(
	s^{{\cal A}\dag}_{i\sigma},
	s^{{\cal B}\dag}_{i\sigma}
\right)$,
$\sigma$ is the spin projection. The Hamiltonian
$H_{\rm E}$ ($H_{\rm A}$) corresponds to graphene (graphane). Below, label
`E' (`A') is used to
denote quantities associated with graphEne (graphAne). The Hamiltonian
$H_{\rm E}$ is the usual graphene Hamiltonian corresponding to
$p_z$-electrons of
carbon hopping between nearest carbon atoms arranged into the honeycomb
lattice. For such lattice, the electron creation operators are arranged
into a spinor $P_{i\sigma}^\dag$, where $i$ denotes the bi-atomic unit cell
of the lattice. The spinor
component labeled
`${\cal A}$'
(`${\cal B}$')
corresponds to a site on the
${\cal A}$
sublattice
(${\cal B}$ sublattice).
The hopping matrix
$\hat T_{ij}$
in the spinor representation in momentum space is
\begin{eqnarray}
\label{t_k}
\nonumber \hat T_{\bf k}&=&\left(
\begin{matrix}
0&t_{\bf k}\cr
t_{\bf k}^{*}&0\cr
\end{matrix}\right),\\
t_{\bf k}&=&t_p\left[1+2\exp\left(\frac{3ik_x
a_0}{2}\right)\cos\left(\frac{\sqrt{3}k_ya_0}{2}\right)
	      \right].
\end{eqnarray}
The Hamiltonian $H_{\rm A}$
is a simplified model of graphane. It describes the $p_z$-electrons of
graphene hybridized with the $s$-electrons of hydrogen, attached to each
carbon atom. Other bands are disregarded. The carbon-hydrogen hybridization
constant
$t_0 =5.8$\,eV
exceeds the carbon-carbon hopping amplitude
$t_p = 2.7$\,eV
and the relative energy of the hydrogen $s$-orbital
$\varepsilon_{\rm H}=0.4$\,eV.~\cite{schmidt2010} 

In ${\bf k}$-space the Hamiltonian 
$H_{\rm A}$
can be expressed as
\begin{eqnarray}
H_{\rm A}
=
\left(
	\begin{matrix}
		\hat T_{\bf k} & t_0 \sigma_0 \\
		t_0 \sigma_0 & \varepsilon_{\rm H} \sigma_0 \\
	\end{matrix}
\right),
\label{H_A_matrix}
\end{eqnarray}
where 
$\sigma_0$
is the 2x2 unity matrix. Here the upper left 2x2 corner corresponds to the
carbons atoms, lower right 2x2 block corresponds to the hydrogens, the
remaining blocks describe the C-H hopping.

The matrix for 
$H_{\rm A}$
is easy to diagonalize. As a result we obtain four graphane bands:
\begin{eqnarray}
\varepsilon^{\rm A}_{m}
=
\frac{1}{2}
\left(
	\pm |t_{\bf k}| \pm \sqrt{4 t_0^2 + |t_{\bf k}|^2}
\right),
\quad
m = 1,2,3,4.
\end{eqnarray}
In this formula 
$\varepsilon_{\rm H}$
is neglected for it is small.

Although only four bands in graphane are considered in our model
Hamiltonian,
Eq.~\eqref{H_A},
it captures the main features of graphane: at half filling,
Eq.~\eqref{H_A}
describes an insulator with a gap located at the $\Gamma$ point. The value
of the gap
$E_g$
(for the parameters written above
$E_g=6.0$\,eV)
is found to be consistent with
Ref.~\onlinecite{sofo_graphane,lebegue_2009}.
Note, however, that there is no consensus about the exact values of the
graphane model parameters. But high precision is not important for the
qualitative results obtained below.

The Hamiltonian $H_{\rm A}$ is valid when the numbers of hydrogen and
carbon atoms are equal. If at
some site hydrogen is absent, then the hydrogen $s$-orbital is not
available for the electrons. This constraint may be enforced by introducing
an infinitely-strong repulsion between the ``hydrogen hole" and the
electron on the $s$-orbital:
\begin{eqnarray} 
\label{Falicov-Kimball}
H_{\rm EA}\!
=\!
H_{\rm A}\!+
U \!\sum_{i\sigma}\!
	S^{ \dag}_{i\sigma}\widehat N^{\rm hh}_i S_{i\sigma},
\\
\widehat N^{\rm hh}_i\!=\!{\rm diag}(n^{\rm hh}_{{\cal A}i},n^{\rm
hh}_{{\cal B}i}),
\end{eqnarray} 
where $U\rightarrow+\infty$, and $n^{\rm hh}_{{\cal A,B}i}$ are the numbers
of hydrogen holes at
site $i$. These numbers can randomly take the values $0$ or $1$ with
mean value $\langle n^{\rm hh}_{{\cal A,B}i}\rangle=n^{\rm hh}$, where
$n^{\rm hh}$ is the
concentration of hydrogen holes per carbon atom.
The Hamiltonian
$H_{\rm EA}$
is a version of the Falicov-Kimball model in which mobile $p$- and
$s$-electrons interact with immobile ``holes" whose
concentration
$n^{\rm hh}$
is fixed externally. Thus,
$n^{\rm hh} = 1$
refers to graphene,
$n^{\rm hh} = 0$
refers to graphane. Below, we will study partial hydrogenation:
$0 < n^{\rm hh} < 1$.

\section{Calculations}
\label{calcs}

An analogy between
$H_{\rm EA}$,
Eq.~(\ref{Falicov-Kimball}),
and the Falicov-Kimball model is very useful for our purposes since the
latter model experiences phase separation in a broad range of
parameters.~\cite{chaika1,chaika2,chaika3,kuras1,kuras2}
The reasons for the existence of phase separation here can be understood
with the help of simple arguments. Since
$t_0$
exceeds
$t_p$,
let us study the limit
\begin{eqnarray}
t_0 \gg t_p.
\label{small_tp}
\end{eqnarray} 
We now introduce the electron operators $a$, $b$ diagonalizing those terms
of
$H_{\rm EA}$ which do not involve the carbon-carbon hopping:
\begin{eqnarray}
\label{local_terms}
\nonumber
H_{\rm EA}-
H_{\rm E}
&=&
\sum_\alpha
	a^\dag_\alpha
	a_\alpha \left[
			t_0( 1 - n^{\rm hh}_\alpha  )
			+
			U n^{\rm hh}_\alpha
		\right]
\\
&&-t_0\sum_\alpha
	b^\dag_\alpha
	b^{\vphantom{\dagger}}_\alpha (1-n^{\rm hh}_\alpha),
\end{eqnarray}
where
\begin{eqnarray}
\nonumber
p_\alpha
&=&
\frac{b_\alpha -a_\alpha }{\sqrt{2}}
(1-n^{\rm hh}_\alpha)
+
n^{\rm hh}_\alpha b_\alpha \,,\\
s_\alpha
&=&
\frac{b_\alpha +a_\alpha }{\sqrt{2}}
(1-n^{\rm hh}_\alpha)
+
n^{\rm hh}_\alpha a_\alpha \,.
\end{eqnarray}
We omit the sublattice and spin labels since the expressions are the
same for any
${\cal A,B}$,
and $\sigma$. The index $\alpha$ labels individual carbon atoms ($i,j$
label unit cells). In
Eq.~\eqref{local_terms}
we neglect the term proportional to
$\varepsilon_{\rm H}$
since $\varepsilon_{\rm H}\ll t_p$ ($\ll t_0$).
It follows from
Eq.~(\ref{local_terms})
that the on-site energy of the fermions $a$ is much higher than the on-site
energy of $b$ for any
$n_{\alpha}^{\rm hh}$,
since
$t_0,U\gg t_p$.
Thus, to lowest order in
$t_p/t_0$,
these states are empty, and can be neglected. In this approximation
\begin{eqnarray}
\nonumber &&H_{\rm EA}
\approx
-t_0\sum_\alpha b^{\dag}_\alpha
		b_\alpha (1-n^{\rm hh}_\alpha)\\
&&-\frac{t_p}{2}
\sum_{\langle \alpha \beta \rangle}
	b^{\dag}_\alpha b_\beta
	[
		1
		+
		\gamma (n^{\rm hh}_\alpha + n^{\rm hh}_\beta)
		+
		\gamma^2n^{\rm hh}_\alpha n^{\rm hh}_\beta
	],
\\
&&\gamma=\sqrt{2}-1 \approx 0.41,
\end{eqnarray}
where
$\langle \ldots \rangle$
denotes summation over the nearest neighbors. From this equation we see
that to separate two hydrogen holes sitting on neighboring sites one must
spend an energy of the order of
$t_p \gamma^2\langle b_\alpha^\dag b_\beta\rangle$.
This corresponds to the \textit{attraction} between the hydrogen holes (and
between
the hydrogen atoms) as in the model used in
Ref.~\onlinecite{shytov_separation}. This attraction induces the phase
separation. The additional correlations between the adsorbed adatoms (e.g.,
due to bond reorganization in graphane),
which our model neglects, may be incorporated as an effective short-range
attraction between the hydrogens. The effect of this attraction is obvious:
it favors phase separation.

Using $H_{\rm EA}$,
Eq.~\eqref{Falicov-Kimball},
we can derive the equation of motion for the single-electron Green's
function in the
$(\omega,\mathbf{k})$
representation:
\begin{eqnarray}
\label{Green}
\nonumber (\omega + \mu ) \hat G_{pp} + \hat T_{\bf k} \hat G_{pp} + t_0
\hat G_{sp}=1,
\\
(\omega + \mu + \varepsilon_{\rm H}) \hat G_{sp} + t_0
\hat G_{pp}-U \hat F_{sp}= 0.
\end{eqnarray}
Here $\mu$ is the chemical potential, $\hat G_{pp,sp}$ and $\hat F_{sp}$
are the Fourier transforms of the time-ordered propagators
\begin{eqnarray}
\nonumber
\hat{G}_{pp}(i-j,t)
&=&
-i\left\langle
	T\,P_{i\sigma}(t)P_{j \sigma}^{\dag}(0)
\right\rangle,
\\
\hat{G}_{sp}(i-j,t)
&=&
-i\left\langle
	T\,S_{i\sigma}(t)P_{j \sigma}^{\dag}(0)
\right\rangle,
\\
\nonumber
\hat{F}_{sp}(i-j,t)
&=&
-i\left\langle
	T\,\widehat N^{\rm hh}_i(t)S_{i\sigma}(t)P_{j \sigma}^{\dag}(0)
\right\rangle.
\end{eqnarray}
The propagator $\hat{F}_{sp}$ requires an additional equation of motion,
which relates
$\hat{F}_{sp}$
with the propagator
\begin{eqnarray}
\hat{F}_{pp}
=
-i\langle
	T\,\widehat N^{\rm hh}_i(t)P_{i \sigma}(t)P_{j \sigma}^\dag (0)
\rangle.
\end{eqnarray} 
To truncate the infinite set of equations for the Green's functions, we
apply the Hubbard-I approximation. It is a simple mean-field scheme
suggested in the seminal
papers~\cite{hubbard1}.
The applicability of Hubbard-I and related approaches has been tested in
many cases (see, e.g.,
Refs.~\onlinecite{hubbard1,beenen,rozhkov_rakhmanov}).
In the Hubbard-I approach,
$\hat{F}_{pp}$
is approximated by the product
\begin{eqnarray}
\hat{F}_{pp}=
\langle \widehat N^{\rm hh}
\rangle\hat{G}_{pp}=n^{\rm hh}\hat{G}_{pp}.
\end{eqnarray} 
This closes the system of equations
\eqref{Green},
whose solution may be written explicitly as
\begin{eqnarray}
\nonumber \hat{G}_{pp}&=&\frac{\omega + \mu + \varepsilon_{\rm H}}
{(\omega + \mu + \varepsilon_{\rm H})(\omega + \mu + \hat{T}_{\bf
k})-n^{\rm H} t_0^2},\\
\hat{G}_{sp}&=&-\frac{n^{\rm H} t_0}{(\omega + \mu + \varepsilon_{\rm
H})(\omega + \mu + \hat T_{\bf k})-n^{\rm H} t_0^2},\label{Gpp_sp}
\end{eqnarray}
where
$n^{\rm H} = 1 - n^{\rm hh}$
is the hydrogen concentration per carbon atom. These equations are obtained
in the limit $U\to\infty$. Similarly, the Green's function
\begin{eqnarray}
\hat{G}_{ss}(i-j,t)=-i\left\langle T\,S_{i\sigma}(t)
S_{j \sigma}^{\dag}(0)\right\rangle
\end{eqnarray} 
is calculated
\begin{eqnarray}\label{Gss}
\hat G_{ss}
&=&
n^{\rm H}\frac{1-t_0 \hat G_{sp}}{\omega + \mu + \varepsilon_{\rm H}}.
\end{eqnarray}
In the limiting case
$n^{\rm H}=0$
($n^{\rm H}=1$),
the Green's functions in Eqs.~(\ref{Gpp_sp}) and (\ref{Gss})
coincide with the exact Green's functions corresponding to the Hamiltonian
$H_{\rm E}$
of graphene
($H_{\rm A}$
of graphane).

\begin{figure}
\centering
\leavevmode
\includegraphics[width=0.99\columnwidth]{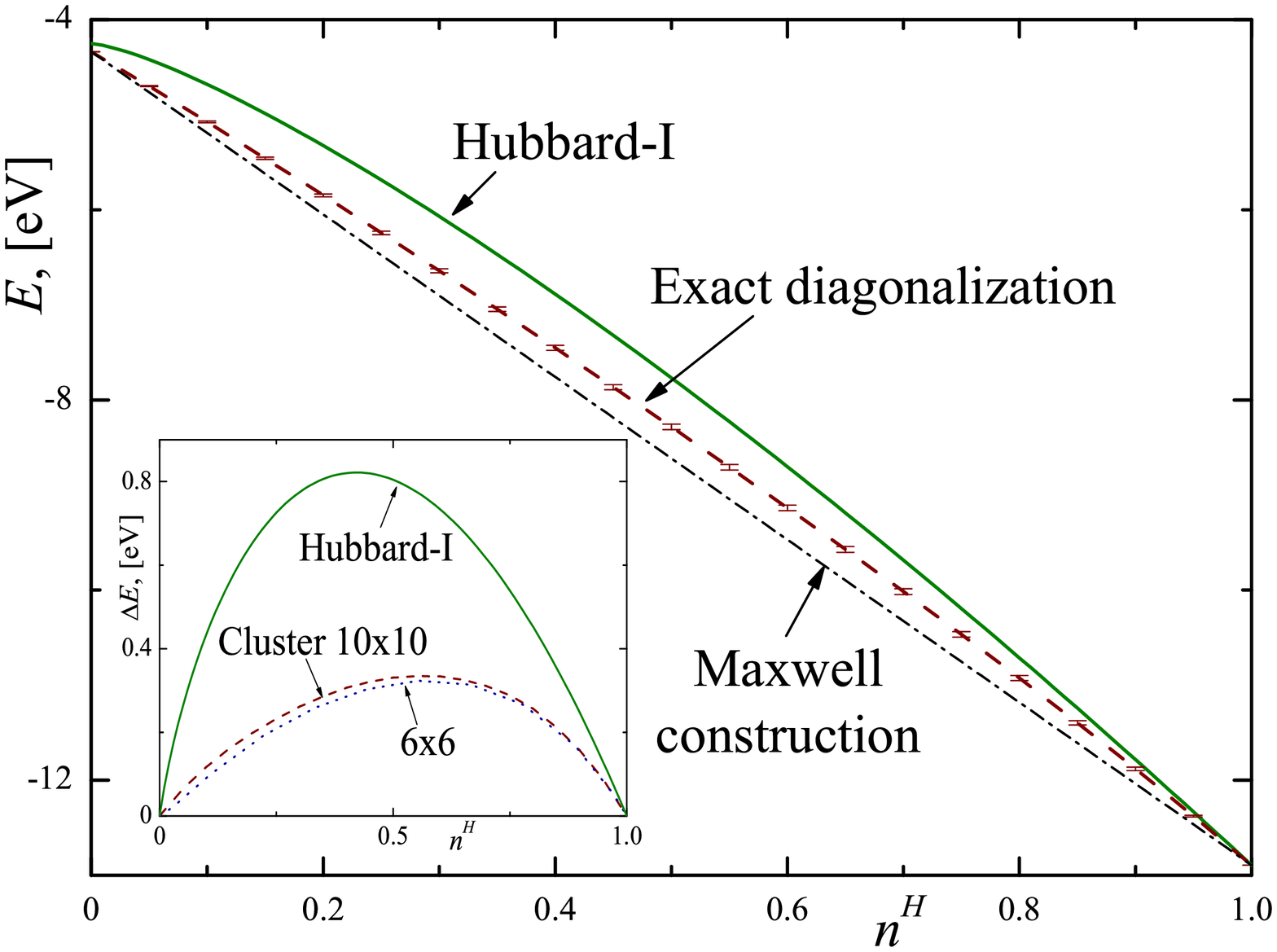}
\caption[]
{(Color online) Electron energy $E$ vs the concentration $n^{\rm H}$ of
hydrogen adatoms calculated in the Hubbard-I approximation (green solid
curve) and by exact diagonalization of $10\times10$ unit cells cluster (red
dashed curve). The negative curvature of $E(n^{\rm H})$ over the whole
range of $n^{\rm H}$ is an indication of the instability of the system
toward the macroscopic separation into phases with $n^{\rm H} = 0$ and
$n^{\rm H} = 1$. The Maxwell construction is shown by blue dot-dashed line.
The inset shows the energy difference between homogeneous and phase
separated states calculated in the Hubbard-I approximation (green solid
curve) and by exact diagonalization of $10\times10$ and $6\times6$ clusters
(red dashed and blue dotted curves, respectively). The model parameters
are: $t_0=5.8$\,eV, $t_p=2.7$\,eV, $\varepsilon_H=0.4$\,eV. For exact
diagonalization, $U=400$\,eV.
}\label{fig::energy}
\end{figure}

When the Green's functions are known, the density of states, the electron
concentration, and the energy can be calculated as a function of $\mu$.
Fixing the electron concentration
$(1 + n^{\rm H})$ per carbon atom, we find $\mu = \mu(n^{\rm H})$ and the
energy
$E = E [\mu(n^{\rm H})]$ at $T = 0$. 

The Hubbard-I results are presented in
Fig.~\ref{fig::energy}.
The energy-versus-density curve has negative curvature for any 
$n^{\rm H}$. 
This indicates the instability of the homogeneous phase toward the phase
separation. The energy of the phase separated state can be found with the
help of the Maxwell construction. In our case, it is simply a straight line
connecting the energy of the pure graphene at 
$n^{\rm H}=0$
and the energy of the fully hydrogenated graphane at
$n^{\rm H}=1$.
This means that the separated phases are pure graphene and pure graphane.

\begin{figure}
\centering
\leavevmode
\includegraphics[width=0.99\columnwidth]{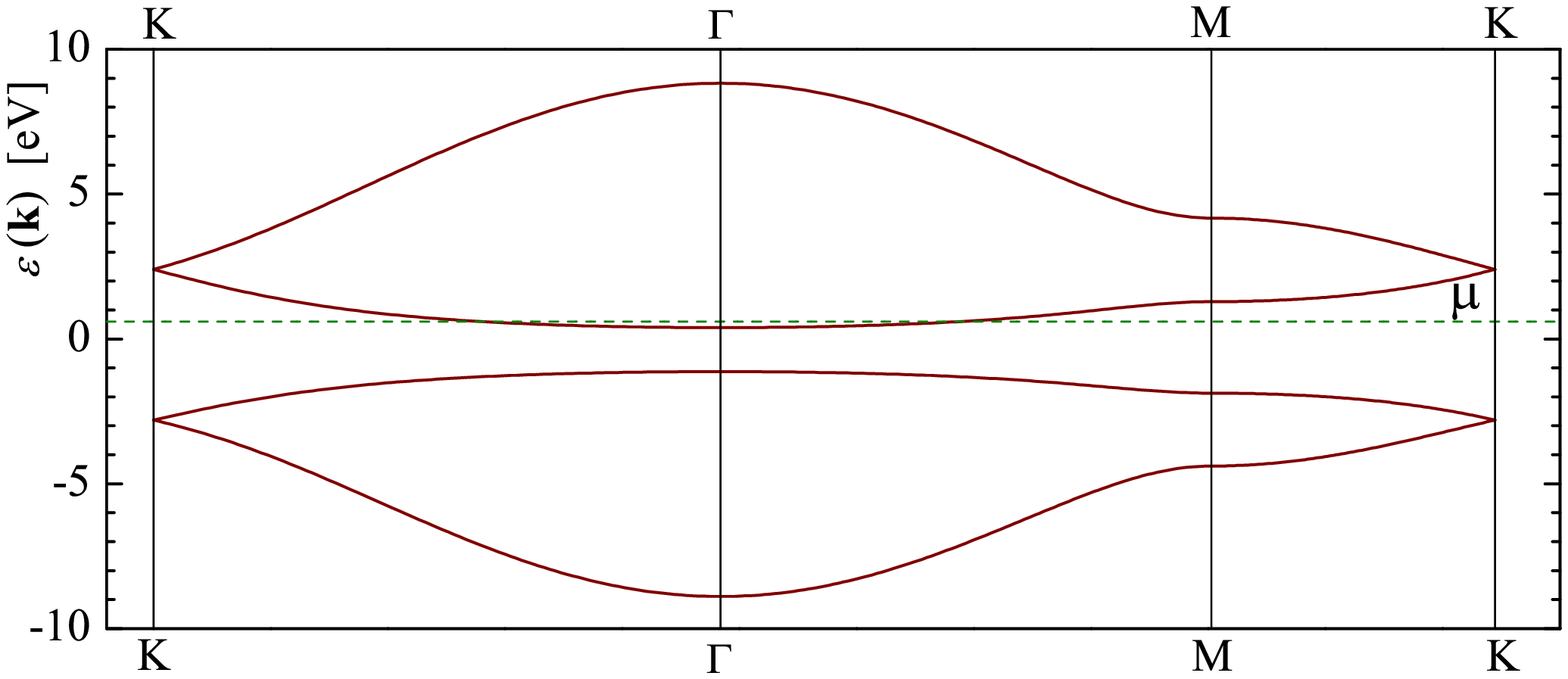}
\caption[]
{(Color online) Electronic dispersion
$\varepsilon ({\bf k})$
of the unstable homogeneous metallic phase for
$n^{\rm H} = 0.2$.
Four single-electron bands, found with the help of
Hubbard-I~\cite{hubbard1,hubbard2} are plotted for
different points of the Brillouin zone. The chemical potential $\mu$ is
marked by the horizontal green dashed line; $\mu$ was calculated
self-consistently to ensure
that the electron concentration is
$(1+n^{\rm H})$
per carbon atom. The gap between the conducting and the valence bands at
$n^{\rm H} = 0.2$
is smaller than the graphane gap
($n^{\rm H} = 1$).
}\label{fig::metal}
\end{figure}

The single-electron band structure of the unstable mixed graphene-graphane
phase is shown in
Fig.~\ref{fig::metal}.
It is interesting to note that the homogeneous phase is a metal. This is
consistent with numerical results for small
clusters.~\cite{roman_clustering2009}

\section{Numerical calculations}
\label{numerics}

To check our analytic approach, we also perform exact
diagonalization of the Hamiltonian~\eqref{Falicov-Kimball} on a finite
honeycomb cluster containing $10\times10$ unit cells ($200$ carbon
atoms). Periodic boundary conditions are used. For each $n^H$, hydrogen
atoms are randomly distributed on the cluster, and we calculate the system
energy by averaging it over $1500$ configurations.

\begin{figure}
\centering
\leavevmode
\includegraphics[width=0.99\columnwidth]{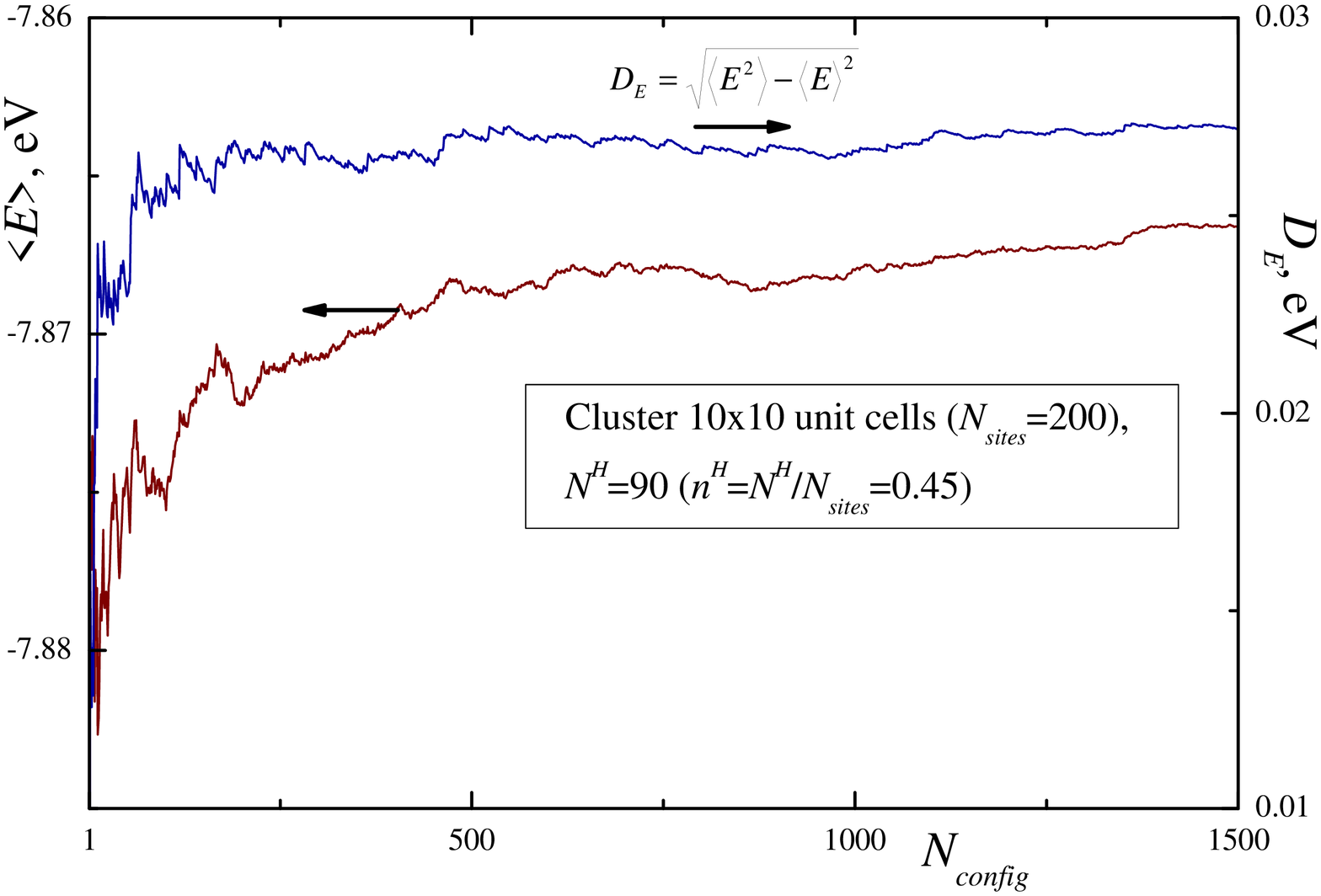}
\caption[]
{(Color online) 
The lower (brown) curve represents the averaged energy 
$\langle E \rangle$,
Eq.~(\ref{disordered_e}).
The upper (blue) curve represents the dispersion,
Eq.~(\ref{dispersion}).
The curves are plotted as functions of the number of the disorder
realizations
$N_{\rm config}$.
Both curves demonstrate saturation for 
$N_{\rm config} \gtrsim 750$.
}\label{fig::size_dep1}
\end{figure}

To check the reliability of the numerical results we investigate their
dependence on the number of sites in the cluster 
($N_{\rm sites}$)
and the number of the configurations used for the averaging
($N_{\rm config}$).
In
Fig.~\ref{fig::size_dep1}
the averaged energy (normalized per site)
\begin{eqnarray}
\langle E \rangle 
=
\frac{1}{N_{\rm config}{N_{\rm sites}}}
\sum_{\Theta=1 }^{N_{\rm config}}  
	E[\Theta]
\label{disordered_e}
\end{eqnarray}
and the normalized energy dispersion
\begin{eqnarray}
\label{dispersion}
D_{\rm E}
=
\sqrt{
	\langle E^2 \rangle
	-
	\langle E \rangle^2
     }
\end{eqnarray}
are plotted as functions of 
$N_{\rm config}$.
In 
Eq.~(\ref{disordered_e})
index $\Theta$ labels different disorder realizations, 
$E[\Theta]$
is the energy for a given disorder realization $\Theta$. Both 
$\langle E \rangle$
and
$D_{\rm E}$
demonstrate saturation for 
$N_{\rm config} \gtrsim 750$.
This suggest that 
$N_{\rm config} = 1500$
we used in our numerical calculations is sufficient to obtain reliable
results.

\begin{figure}
\centering
\leavevmode
\includegraphics[width=0.99\columnwidth]{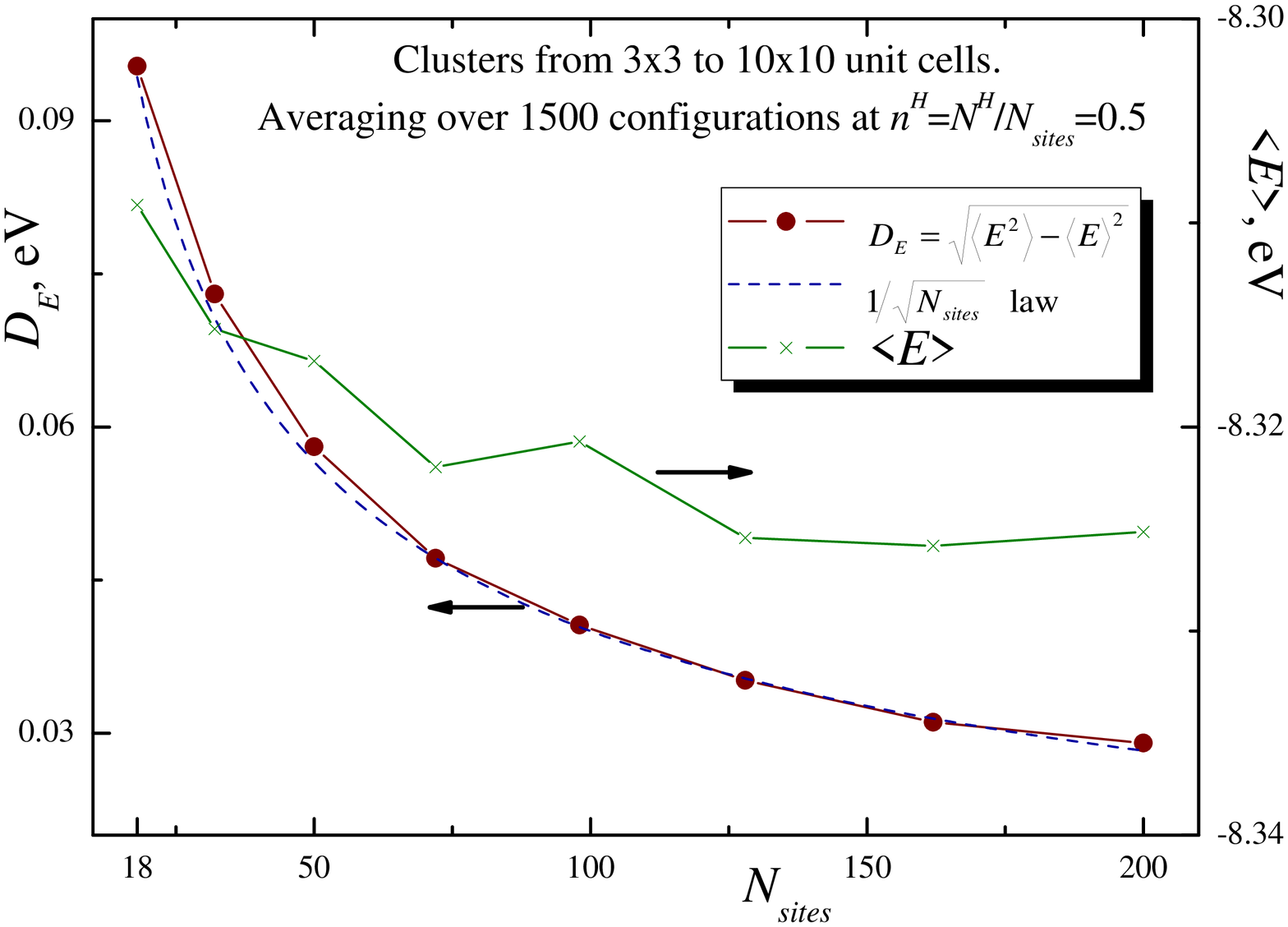}
\caption[]
{(Color online) 
Finite size effects. The solid (green) curve connecting skew crosses
represents the averaged energy 
$\langle E \rangle$,
Eq.~(\ref{disordered_e}).
It saturates for
$N_{\rm sites} \gtrsim 125$.
The solid (red) curve connecting filled circles represents the energy
dispersion,
Eq.~(\ref{dispersion}).
It decays as
$N_{\rm sites}^{-1/2}$.
The decay of the dispersion implies that for large samples the energy is a
self-averaging quantity. The value of the dispersion may be used to
evaluate the accuracy of the estimated value of the energy. For 
$N_{\rm sites}=200$
the error for 
$\langle E \rangle$
is a fraction of a percent.
}\label{fig::size_dep2}
\end{figure}

In 
Fig.~\ref{fig::size_dep2}
the same quantities are plotted versus
$N_{\rm sites}$.
The dispersion decays as
$N_{\rm sites}^{-1/2}$.
This means that the relative strength of the energy fluctuations decreases
when the cluster size grows, and the energy experiences the self-averaging.
The energy itself saturates for
$N_{\rm sites} \gtrsim 125$.
Therefore, our choice of
$N_{\rm sites} = 200$
is adequate. In addition, the ratio
$D_{\rm E}/\langle E \rangle$
sets the relative error for 
$\langle E \rangle$.
For 
$N_{\rm sites} = 200$
this error is a fraction of a percent. We conclude that our numerical
calculations are reliable.

The most important results are shown in
Fig.~\ref{fig::energy}:
observe the negative curvature of the function
$E(n^{\rm H})$.
It implies that the system is unstable and phase separates in two phases:
with
$n^{\rm H} = 0$
(graphene) and with
$n^{\rm H}=1$
(graphane).
Unless
$n^H$
is close to 0 or 1, the energy gain due to the phase separation is of the
order of $10^3$\,K,
see the inset of Fig.~\ref{fig::energy}. Thus,
even at room temperature we can safely use the results obtained at
zero temperature. 

Further, the numerically evaluated energy is of the same order as the
Hubbard-I energy: the magnitude of the Hubbard-I energy is approximately
two times higher than the numerical estimate (see inset in
Fig.~\ref{fig::energy}).
Thus, the qualitative consistency between the numerical calculations and
the Hubbard-I results provides firm support to the findings of
Sec.~\ref{calcs}.

\section{Interface tension and interface stability}
\label{interface}

In the phase-separated state there is a boundary between graphene and
graphane. The geometry of the stable inhomogeneous state depends on the
sign
and the value of the interface tension $\sigma_{0}$. If $\sigma_{0}<0$,
then the
inhomogeneous phase breaks into small clusters to maximize the boundary
length. In the case of $\sigma_{0}>0$, the interface tension acts to
minimize
the length of the graphene-graphane border. In the case of a long strip
this border is a straight line (if the concentration of the hydrogen
adatoms is not small). However, at finite temperatures, even for positive
$\sigma_{0}$, small thermal fluctuations destroy the perfect smoothness of
the boundary between the two phases. The difference in the lattice
symmetry between graphene and graphane, at the level of the electron model,
manifests itself through the values of the orbital overlaps. In the model
considered here we make an approximation regarding the orbital overlaps: we
assume that several of them are equal to zero. 

Further, we neglect the difference between lattice constants in graphene
and graphane. The contribution of the electron-electron interaction to the
interface tension is also disregarded (we briefly discuss the effect of the
interaction below). In other words, the interface tension in our model
arises only due to the electron motion through the graphene-graphane
boundary. 

These assumptions can be justified {\it post factum}: (i) from our model it
follows that the binding energy between a hydrogen atom and the
graphene-graphane interface is of the order of $t_p$, which is consistent
with the results presented in
Ref.~\onlinecite{interface_stability2010};
(ii) we pointed out above that the value of the graphane gap in our
simplified model
Eq.~(\ref{H_A})
turns out to be consistent with other studies; (iii) we found that for
intermediate hydrogenation the stable homogeneous phase is a metal, in
agreement with
Ref.~\onlinecite{roman_clustering2009}.

We will now evaluate
$\sigma_{0}$ 
in the limit
Eq.~(\ref{small_tp}).
In this approximation, electrons in graphane are localized on the C-H
valence bonds [see
Eq.~(\ref{local_terms})]
and their contribution to $\sigma_{0}$ is small (this contribution is
proportional to
$t_p^2/t_0$
and
$\varepsilon_{\rm H}$).
In graphene, electrons are moving from one carbon atom to its nearest
neighbors. However, the electrons from graphene cannot penetrate into
graphane since they have to overcome the graphane gap, which, 
according to
Eq.~(\ref{local_terms}),
is of the order of
$t_0$
when
Eq.~(\ref{small_tp})
holds. Thus, each carbon-carbon bond connecting an atom in graphene with an
atom in graphane does not contribute to the graphene electron kinetic
energy. This, in effect, is equivalent to an increase in the kinetic energy
of the electrons in graphene. The longer the interface, the larger the
number of ``broken" bonds. Thus,
\begin{eqnarray}
\sigma_{0}\sim
	\frac{\kappa\,\varepsilon_{\rm b}}{a_0},
\end{eqnarray} 
where
$\varepsilon_{\rm b}$
is the kinetic energy for each carbon-carbon bond. The numerical
coefficient
\begin{eqnarray}
\kappa = 
	\left\{
	\begin{array}{c}
		1/\sqrt{3} \approx 0.6 {\rm \ for\ zigzag}\\
		2/{3} \approx 0.7 {\rm \ for\ armchair}\\
	\end{array}
 	\right.
\end{eqnarray} 
characterizes the linear concentration of the carbons on the interface.
The kinetic energy per bond is equal to
\begin{eqnarray}
\varepsilon_{\rm b}
=
\frac{2}{3}
\int \!\! d^2 {\bf k} \; \frac{S_0|t_{\bf k}|}{(2\pi)^2}
\; \sim \; t_p\,,
\end{eqnarray}
where the integration is performed over the first Brillouin zone,
$S_0 = 3 \sqrt{3} a^2_0/2$
is the area of the graphene unit cell, the factor 2 corresponds to two
spin projections, (1/3) enters since there are three bonds in a graphene
unit cell, and
$t_{\bf k}$
is defined by
Eq.~(\ref{t_k}). After integration, we have
$\varepsilon_{\rm b}\approx 1.05\,t_p$,
and in our approximation,
\begin{eqnarray}
\sigma_{0}\sim 0.6\,t_p/a_0.
\end{eqnarray} 
A more accurate calculation (following
Ref.~\onlinecite{balian_bloch})
provides
\begin{eqnarray} 
\sigma_{0} \approx 0.2 t_p/a_0\approx 0.6\,{\rm eV}/a_0.
\end{eqnarray} 
In the calculations presented above the contribution of the
electron-electron interaction to
$\sigma_0$
is disregarded. The detailed account of the interaction goes beyond
the scope of the present study. Yet, we would like to offer two
observations. First, the contribution due to the interaction is of the same
order as
$\sigma_0$.
Indeed, the latter originates mostly from the energy of C-C bond. The
chemical energy of C-H bond is of the same order (few eV). Thus, there is
no energy scale in the system which would be able to generate an
overwhelmingly large contribution to the interface tension. Second, the
contribution due to the interaction increases the tension. To prove this,
let us neglect the interaction in the bulk, as it is usually done for
graphene, but retain the interaction term for the electrons near the
graphene-graphane edge. This assumption mimics relative importance of the
interactions for electrons in lower dimensions. It is known that the
repulsive interaction gives positive contribution to the electron energy
(see, e.g., Sec.\,I, \S\,6 of 
Ref.~\onlinecite{landau_stat_phys_p2})
and, consequently, to the interface tension.

We neglect the effect of the temperature $T$ on the phase separation since
the characteristic energies of the problem are much higher than $k_BT$
for any realistic $T$. However, the temperature fluctuations could affect
the smoothness of the graphene-graphane interface even under such
conditions. Following
Ref.~\onlinecite{Chaikin},
we can express the average square fluctuation of the deviation $u$ of the
interface having a length
$L$ as 
\begin{eqnarray}
\langle u^2\rangle= \frac{k_BTL}{2\pi\sigma_{0}}.
\end{eqnarray} 
Thus, we obtain
\begin{eqnarray}
\frac{\langle u^2\rangle}{a^2_0}
\approx 
\left(
	\frac{L}{a_0}
\right)
\left(
	\frac{k_BT}{t_p}
\right).
\end{eqnarray} 
Using the value of the carbon-carbon hopping
$t_p=2.8$~eV,
we find that, at room temperature, the graphene-graphane interface
remains atomically-flat
($\langle u^2\rangle/a^2_0\leq 1$)
over distances
\begin{eqnarray}
L_1\approx 100\,a_0.
\end{eqnarray} 
Note that the estimated values of $\sigma_{0}$ and, consequently, $L_1$
will be larger if one takes into account the contribution to the interface
tension due to the difference between lattice constants in graphene and
graphane.


\section{Conclusion} 
\label{conclusions}

We mapped the model of hydrogen atoms adsorbed on
graphene on a Falicov-Kimball-like model. We demonstrated that this system
has a strong tendency to phase separate. The thermodynamically stable state
is inhomogeneous: all adatoms cluster together, forming two phases:
hydrogen-saturated graphane and hydrogen-free graphene. The interface
between these phases has finite and positive interface tension, which means
that the boundary is stable and flat (if the number of hydrogen adatoms is
not small). The estimated value of the interface tension is high and, at
room temperature, the interface remains atomically flat over distances of
about
$10^2$
lattice constants. This result may be of interest for fabricating graphene
mesoscopic devices with weak edge scattering.

\section{Acknowledgements}

We would like to thank L.~Openov for discussions and suggestions.
This work was supported in part by JSPS-RFBR Grant
No.~09-02-92114 and RFBR Grant No.~09-02-00248.
FN was partially supported by LPS, NSA, ARO, NSF grant No.~0726909,
Grant-in-Aid for Scientific Research~(S), MEXT Kakenhi on Quantum
Cybernetics, and the JSPS via its FIRST program. 
AOS acknowledges partial support from the Dynasty Foundation.

\bibliographystyle{apsrev_no_issn_url}
\bibliography{graphane3}

\end{document}